\documentclass{article}

\usepackage{arxiv}

\usepackage[utf8]{inputenc} 
\usepackage[T1]{fontenc}    
\usepackage{hyperref}       
\usepackage{url}            
\usepackage{booktabs}       
\usepackage{amsfonts}       
\usepackage{nicefrac}       
\usepackage{microtype}      
\usepackage{lipsum}		
\usepackage{graphicx}
\usepackage{natbib}
\usepackage{doi}

\title{Research evolution of metal organic frameworks: A scientometric approach with human-in-the-loop}

\author{Xintong Zhao\\
	Metadata Research Center\\
	Drexel University\\
	Philadelphia, PA, 19104 \\
        \texttt{xintong.zhao@drexel.edu}\\
	\And
	Kyle Langlois \\
        Department of Chemistry\\
	University of Central Florida\\
	Orlando, FL, 32816\\
	\And
	Jacob Furst \\
        Department of Chemistry\\
	University of Central Florida\\
	Orlando, FL, 32816\\
	\And
	Yuan An \\
	Metadata Research Center\\
	Drexel University\\
	Philadelphia, PA, 19104 \\
	\And
	Xiaohua Hu \\
	Metadata Research Center\\
	Drexel University\\
	Philadelphia, PA, 19104 \\
	\And
	Diego Gomez Gualdron \\
        Department of Chemical \& Biological Engineering\\
	Colorado School of Mines\\
	Golden, CO, 80401\\
	\And
	Fernando Uribe-Romo \\
        Department of Chemistry\\
	University of Central Florida\\
	Orlando, FL, 32816\\
	\And
	Jane Greenberg \\
	Metadata Research Center\\
	Drexel University\\
	Philadelphia, PA, 19104 \\
}

\date{}

\renewcommand{\shorttitle}{\textit{arXiv} Template}

\begin{document}
\maketitle

\begin{abstract}
\textbf{Purpose} This research presents a scientometric analysis bolstered by human-in-the-loop, domain experts, to examine the field of metal-organic frameworks (MOFs) research. Scientometric analyses reveal the intellectual landscape of a field. The study engaged MOF scientists in the design and review of our research workflow. MOF materials are an essential component in next-generation renewable energy storage and biomedical technologies. The research approach demonstrates how engaging experts, via human-in-the-loop processes, can help develop a comprehensive view of a field’s research trends, influential works, and specialized topics.

\textbf{Design/Methodology/Approach} A scientometric analysis was conducted, integrating natural language processing (NLP), topic modeling, and network analysis methods. The analytical approach was enhanced through a human-in-the-loop iterative process involving MOF research scientists at selected intervals. MOF researcher feedback was incorporated into our method. The data sample included 65209 MOF research articles. Python3 and software tool VOSviewer were used to perform the analysis.

\textbf{Findings} The findings demonstrate the value of including domain experts in research workflows, refinement, and interpretation of results. At each stage of the analysis, the MOF researchers contributed to interpreting the results and method refinements targeting our focus on MOF research. This study identified influential works and their themes. Our findings also underscore four main MOF research directions and applications.

\textbf{Research Limitations} This study is limited by the sample (articles identified and referenced by the Cambridge Structural Database) that informed our analysis.

\textbf{Practical Implications} Our findings contribute to addressing the current gap in fully mapping out the comprehensive landscape of MOF research. Additionally, the results will help domain scientists target future research directions.

\textbf{Originality/Value} To the best of our knowledge, the number of publications collected for analysis exceeds those of previous studies. This enabled us to explore a more extensive body of MOF research compared to previous studies. Another contribution of our work is the iterative engagement of domain scientists, who brought in-depth, expert interpretation to the data analysis, helping hone the study.

\end{abstract}


\section{Introduction}

Life sciences and environmental studies have impacted nearly every sectors of society, with many studies having long-term implications. Recent advancements in environmental science, biomedicine, and computing techniques have motivated an evolving cross-disciplinary field known as "metal-organic frameworks (MOFs)" has garnered significant interest. Due to their unique material characteristics, MOFs materials are often considered as a platform for various next-generation applications, such as renewable energy storage, carbon capture, drug delivery and disease diagnosis, among other applications\cite{wang2016applications,cavka2008new,cote2005porous}.

An emerging field, MOFs research developed by integrating  a range of synthesis disciplines\cite{kinoshita1959crystal,ginsberg1990early,moulton2001molecules} into what now known as Reticular Chemistry\cite{yaghi2019introduction}. Since the discovery of the first MOF discovery in 1999 \cite{li1999design}, research on MOF materials has had a continuous upward trend. Moreover, MOFs research has rapidly  expanded into an array of real-life applications.

The wide-spread increase in MOF publications reveal a diversity of scientific approaches and research foci.  As a result, researchers faces in manually assessing the vast amount of information associated with this growing discipline. Researchers seek a concise approach to comprehend and engage in this field of study. This is particularly true for new researchers, who need to understand the historical and developing research landscape. 

Scientometrics, the “quantitative study of science, communication in science, and science policy” \cite{sengupta1992bibliometrics,vinker1991possible,hess1997science}, hold a distinct advantage in tackling the mentioned challenge and researchers' needs. The ability to analyze large amounts of scholarly data, identify research trends and latent connections between disparate research topics \cite{04817c26-80f1-38c3-8cf5-03c7fbde591b,Sebastian_Siew_Orimaye_2017} positions scientometric methods as a possible solution. Due to its effectiveness, numerous scientometric analyses were previously conducted in economics and business studies\cite{castillo2018bibliometric,merigo2017bibliometric,bonilla2015economics,ye2020bibliometric}, social studies\cite{de2005bibliometric,guo2019bibliometric}, and especially biomedical studies\cite{guo2020artificial,krishnamoorthy2009bibliometric}: for example, Song et al.\cite{Jeong_Xie_Yan_Song_2020} conducted a literature-based network analysis to examine the drugs and side effects across nearly 170000 publications; Pichika et al.\cite{mak2022success} explored current landscape of AI-driven drug candidates with perspecives for future development. 

Scientometric analysis has been pursued in various  material science areas\cite{ho2014bibliometric,lan2022depth,dong2012bibliometric}, including MOF research. For example, Naseer et al.\cite{naseer2022metal} collected 1187 publications and applied counting-based approaches in bibliometric analysis to explore the research trends and intellectual structure in the direction of wastewater decontamination. Ye and Yang\cite{ye2022exploring} collected 2353 publications and analyzed MOFs publications discussing the use of MOFs in electrochemistry. Shidiq\cite{shidiq2023bibliometric} analyzed 1000 publications in the subarea of nano metal-organic frameworks in medical science to discover occurring phenomena and research themes. And another example is found with  Wang and Ho\cite{wang2016research}, who explored the landscape of MOFs in general by extracting terms from title, author index and abstract of collected articles then manually performed the analysis on Microsoft Excel spreadsheet.

These previous investigations offer important insight into the MOF field, although they have several limitations: 1) The majority of the research  focuses on a narrow topic of MOF research, so the complete landscape of MOF studies remains unclear; 2) Several of the  previously performed analysis are based on small samples and stand as preliminary work, 3) A number of the analyses are conducted manually, relying solely on either human expertise or data perspectives.

Compared to previous related research, our investigation aims to provide a more in-depth analysis and comprehensive view of the MOF community. To meet this research aim, we developed a significantly larger corpus of publication data: our publication data, including citation, article metadata and text, are drawn from 65209 articles, which is more than ten-folds larger than previous studies. We also pursued additional analysis to reveal yet to be explored specific research groups and their topics by applying network analysis and natural language processing techniques. Furthermore, a notable contribution of this study is that we integrated points of view from both data and human experts. Our method addresses a limitation in scientometric analysis \cite{lund2021select}. Many previous studies are either purely based on data or human expertise, which can lead to a range of limitations. Without human expertise, valuable insights and links between different studies that are not apparent to data can be overlooked \cite{zanzotto2019human}, and the latter can be limited by relying on human expertise alone (especially when the domain is multidisciplinary). Hence, we designed our data analysis workflow with human in the loop to deliver a comprehensive landscape of MOFs research: we first conducted data-driven exploratory analysis using scientometric methods, then refined the data analysis by iterative engagement with domain expert scientists. 

The four research objectives are to:

\begin{enumerate}
  \item Identify the most influential research works, journals as well as their latent connections in the area of metal-organic framework materials,
  \item Determine major research directions in MOF area, as well as their trends over time,
  \item Reveal specific communities that exist in the research network, and
  \item Recognize their detailed research topics.
\end{enumerate}

In addition to the above research objectives above, we explored the following question: \textit{Is the most cited study always the most impactful study}? The paper sections that follow present our methods, findings, and discusses the research implications. 

\section{Method}

\begin{figure}[h!]
 \centering
 \includegraphics[height=6.5cm]{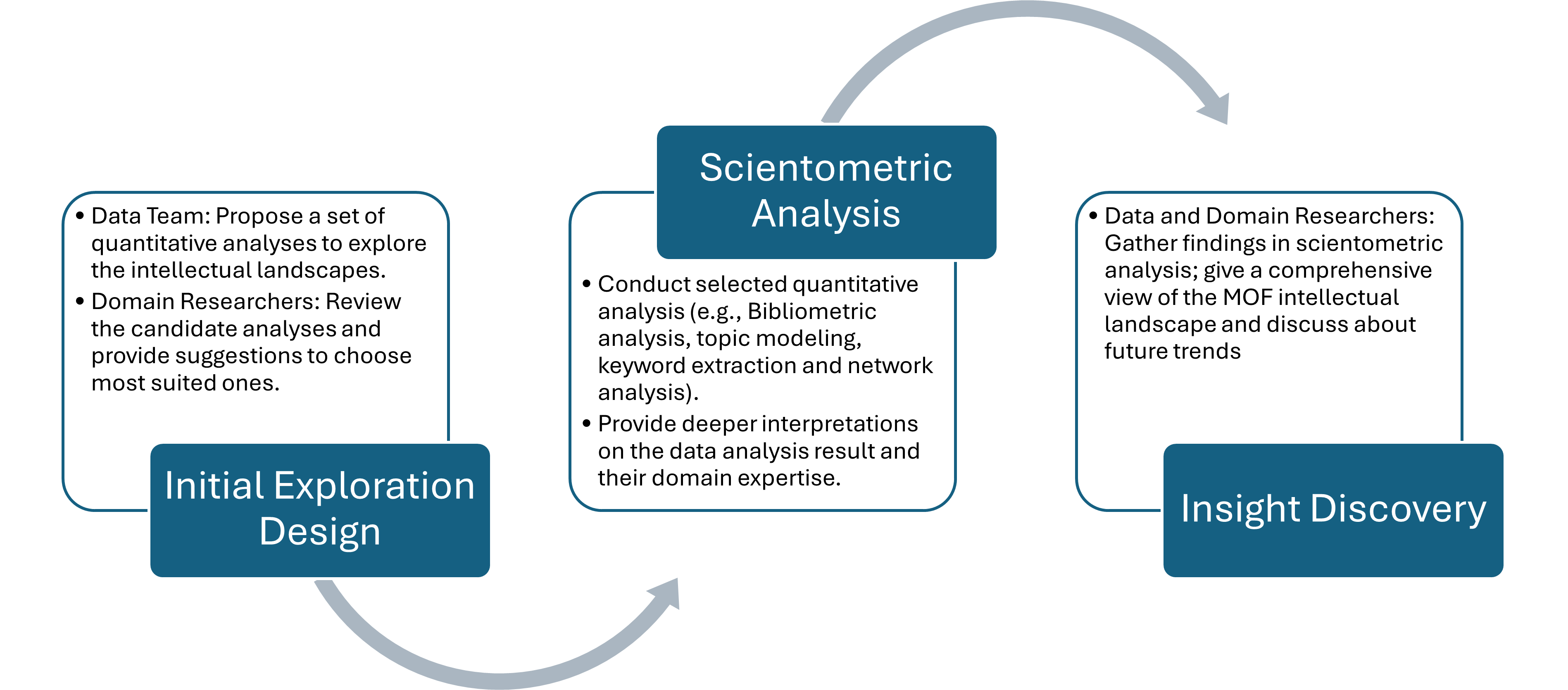}
 \caption{Overall Workflow of the Research Design}
 \label{fig0}
\end{figure}

To answer the research questions outlined in the Introduction, we designed our expert knowledge guided scientometric analysis for discovering the landscape of MOF community in three main steps: 1) data collection and processing, 2) bibliometric and network analysis, 3) specific groups detection and topic modeling. Our domain collaborators, who are scientists in the field of metal-organic frameworks, were involved in each phase of the analysis as illustrated in figure \ref{fig0}.

First the target articles for data collection were validated by multiple scientists. Second, we confirmed that the scientometric research objective met the domain researchers' needs. Third, we pursued the initial scientometric data analysis. Finally, the initial results were assessed and refined by domain experts in two ways: the experts enhanced the initial results by adding insights and latent connections that are not apparent in the data , and provided suggestions to remove analysis components that are not helpful to domain researchers. Specifically, the list of organizations conducting different research topics was removed from the analysis, since researchers' affiliations may change.

\subsection{Data Collection and Processing}

Our data collection and processing process consists of two steps. First, we gather raw data related to MOF scientific publications. This data in raw form may consist of information about citation dependency, abstract and metadata from CrossRef (https://www.crossref.org) and Scopus (https://www.scopus.com/home.uri) indexing systems. Secondly, we parse this raw data to a structured format. From this structured database, we then construct a research graph network for MOFs.

\begin{figure*}
 \centering
 \includegraphics[height=8cm]{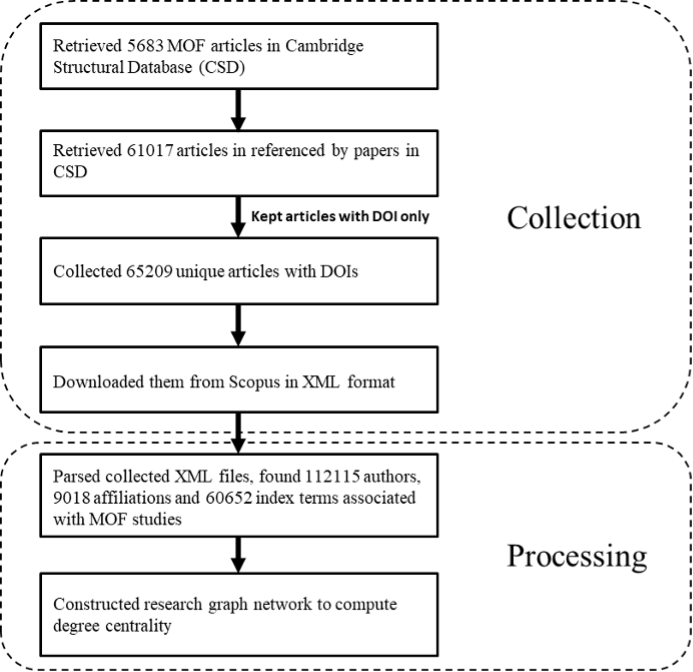}
 \caption{Overall Workflow of Data Collection and Processing}
 \label{fig1}
\end{figure*}

The data collection process was initiated by gathering MOF crystal structures included in the Cambridge Structural Database (CSD)\cite{csd}, which is a database validated by materials scientists containing a wide range of crystal material structures. Our search strictly focused on the MOFCSD database \cite{Moghadam_Li_Wiggin_Tao_Maloney_Wood_Ward_Fairen-Jimenez_2017}, which is a subset of the CSD that contain al “true MOFs.” Materials from related field (e.g., 1-d coordination polymers) are not included. We used Digital Object Identifier (DOI) to identify publications associated with each collected MOF structure. Among 10,636 crystal structures in the database, 5,683 unique scientific articles along with their DOI were identified. For each identified publication, we also retrieved DOIs of papers in the reference list (in JSON - a structured text format) using CrossRef API. In total, we found 65209 unique articles. After gathering a complete list of DOIs for articles above, we used these DOI to retrieve and download the metadata (e.g., title, journal, publication year, keyword, authors, affiliation) along with the abstract of each identified MOF article from Scopus in XML format. After parsing XML files collected from Scopus, we found 112,115 unique authors, 9,018 unique affiliations and 60,652 unique indexed keywords closely related to the metal-organic framework studies. We linked the above metadata to their corresponding articles. An overall workflow of this data collection process is illustrated in Figure \ref{fig1}.

\subsection{MOF Research Network Analysis}

We analyzed MOF research networks from different angles: 1) impactful research entities, 2) change in topics over time and 3) different research communities in the MOF area. The network analysis is performed by Python3 library Networkx\cite{hagberg2008exploring} and software tool VOSviewer\cite{van2010software}.

\subsubsection{Most Impactful Entities in the Field}

In this study, impactful publications include works that are significantly cited by other studies in the same research field. We constructed the MOF research network using Networkx as a directed graph, where each node represents one research publication, edges represent citation dependency. To discover entities (e.g., articles, journals) that produced impactful research output, we used degree centrality scores to quantify the article impact. In Networkx, the degree centrality of a node V is defined as the number of nodes it connects to (both in and out), then normalized by dividing the maximum possible degree of the network \cite{hagberg2008exploring}. 

Based on this definition, a high degree centrality score of a publication indicates a significant connection to other research works. We computed the centrality score for every node in our MOF graph network and sorted scores in descending order. We also list the corresponding number of references of nodes as a comparison.

\subsubsection{Specific Community Detection}

Communities, also called clusters in a graph network, consist of a set of nodes with similar features. The community detection (“clustering”) algorithm used in this paper is developed from a modularity-based algorithm\cite{clauset2004finding} and is implemented by the software VOSviewer\cite{van2010software,clauset2004finding,waltman2010unified}.

To find out specific research groups in the field of MOFs, we conducted network co-citation analysis, which determines closeness of two publications by the number of times that a third publication cites them together.

\subsection{Topic Trends and Modeling}

As part of our analysis, we pursued text analysis, enhanced by domain experts, to detect clusters representing (1) research topic trends over time and (2) detailed research topics in specific research directions. The former part of analysis is done based on the occurrence of indexed terms and implemented by VOSviewer; terms, both unigram terms and n-gram terms are taken into account. The latter part of analysis is conducted using a noise-reducing topic modeling algorithm\cite{churchill2021topic} developed based on Latent Dirichlet Allocation (LDA). For each detected specific community, we gathered abstracts of papers belonging to this community, then applied the topic modeling algorithm on the set of abstracts. Candidate topics generated by the algorithm are analyzed by both data and domain experts.

\section{Results and Findings}

\begin{figure}[h!]
 \centering
 \includegraphics[height=6.5cm]{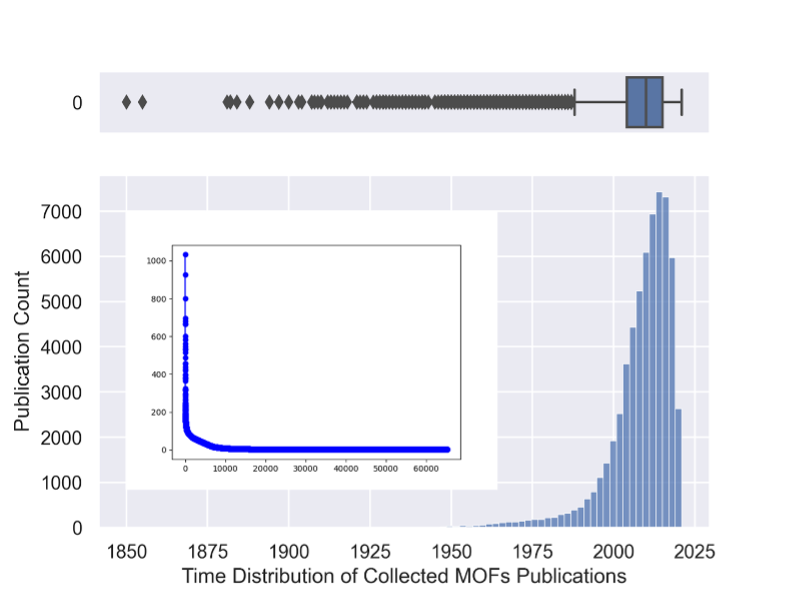}
 \caption{Publication Data Exploration in Metal-Organic Framework Area}
 \label{fig2}
\end{figure}

As an initial data exploration, we plotted the publication distribution by year and the citation-rank distribution in Figure \ref{fig2}, to see the publication and citation pattern.

The publication distribution resembles a Zipf distribution function \cite{egghe2005power,perc2010zipf} with a maximum number of publications peaking in 2013 (near 7k publications) Since approximately 2000, the number of MOF publications started to grow exponentially, with a significant amount of research works being published–75\% of articles between 2005 to 2020. Pre-2000 publications refer mostly to “coordination polymers.” The term “MOF” emerged in 1999, and refers to those coordination polymers that retain a permanently open porous architecture even after guest/solvent removal\cite{li1999design}. Among these publications, their citation-rank distribution seems to be extremely skewed to the right (Fig 2, inset)—this indicates that only a small percentage of publications receive most citations, whereas many publications remain less cited, consistent with a Pareto distribution. The next questions are: what are the characteristics of these highly cited papers? Are these the most influential in the domain science?

\subsection{Most Impactful Entities}

\begin{table*}[]
\small
\caption{\ Top 20 Most Influential Publications based on Degree Centrality}
  \label{tab1}
  \resizebox{\columnwidth}{!}{%
\begin{tabular}{llllll}
\textbf{Publication Title} & \textbf{Journal Name} & \textbf{Year} & \textbf{Type} & \textbf{Referenced By} & \textbf{Centrality} \\
\hline
Single-crystal structure validation with the program PLATON \cite{spek2003single} & Journal of Applied Crystallography & 2003 & Article & 16311 & 0.0158 \\
Functional porous coordination polymers \cite{kitagawa2004functional} & \begin{tabular}[c]{@{}l@{}}Angewandte Chemie -\\ International Ed.\end{tabular} & 2004 & Review & 9724 & 0.0142 \\
A short history of SHELX\cite{sheldrick2008short} & \begin{tabular}[c]{@{}l@{}}Acta Crystallographica Section A:\\  Foundations of Crystallography\end{tabular} & 2008 & Review & 80558 & 0.0123 \\
Reticular synthesis and the design of new materials\cite{yaghi2003reticular} & Nature & 2003 & Review & 8012 & 0.0107 \\
\begin{tabular}[c]{@{}l@{}}Systematic design of pore size and functionality in isoreticular \\ MOFs and their application in methane storage\cite{eddaoudi2002systematic}\end{tabular} & Science & 2002 & Article & 6717 & 0.0104 \\
Luminescent functional metal-organic frameworks\cite{cui2012luminescent} & Chemical Reviews & 2012 & Review & 4960 & 0.0102 \\
\begin{tabular}[c]{@{}l@{}}Selective gas adsorption and separation in metal-organic\\  frameworks\cite{li2009selective}\end{tabular} & Chemical Society Reviews & 2009 & Article & 6936 & 0.0102 \\
Carbon dioxide capture in metal-organic frameworks\cite{sumida2012carbon} & Chemical Reviews & 2012 & Article & 5090 & 0.0092 \\
Metal-organic frameworks for separations\cite{li2012metal} & Chemical Reviews & 2012 & Review & 5284 & 0.0089 \\
Luminescent metal-organic frameworks\cite{allendorf2009luminescent}& Chemical Society Reviews & 2009 & Article & 4407 & 0.0086 \\
Metal-organic framework materials as catalysts\cite{lee2009metal} & Chemical Society Reviews & 2009 & Article & 6826 & 0.0083 \\
Metal-organic framework materials as chemical sensors\cite{kreno2012metal} & Chemical Reviews & 2012 & Review & 5692 & 0.0082 \\
\begin{tabular}[c]{@{}l@{}}Modular chemistry: Secondary building units as a basis for the\\ design of highly porous and robust metal-organic carboxylate\\ frameworks\cite{eddaoudi2001modular}\end{tabular} & Accounts of Chemical Research & 2001 & Article & 5007 & 0.0081 \\
\begin{tabular}[c]{@{}l@{}}From molecules to crystal engineering: Supramolecular \\ isomerism and polymorphism in network solids\cite{moulton2001molecules}\end{tabular} & Chemical Reviews & 2001 & Review & 6437 & 0.0079 \\
Hybrid porous solids: Past, present, future\cite{ferey2008hybrid} & Chemical Society Reviews & 2008 & Article & 5139 & 0.0075 \\
Interpenetrating Nets: Ordered, Periodic Entanglement\cite{batten1998interpenetrating} & Angew Chem Int Ed Engl & 1998 & Review & 3859 & 0.007 \\
Hydrogen storage in metal-organic frameworks\cite{murray2009hydrogen} & Chemical Society Reviews & 2009 & Article & 3954 & 0.007 \\
The chemistry and applications of metal-organic frameworks\cite{furukawa2013chemistry} & Science & 2013 & Review & 9023 & 0.0067 \\
\begin{tabular}[c]{@{}l@{}}A homochiral metal-organic porous material for\\ enantioselective separation and catalysis\cite{seo2000homochiral}\end{tabular} & Nature & 2000 & Article & 3730 & 0.0065 \\
\begin{tabular}[c]{@{}l@{}}Design and synthesis of an exceptionally stable and highly \\ porous metal- organic framework\cite{li1999design}\end{tabular} & Nature & 1999 & Article & 6486 & 0.0064\\
\hline
\end{tabular}
}
\end{table*}

To identify influential MOF research work, we computed the centrality value of each node inside the constructed research network (each node represents one scientific publication in MOFs area); we sorted the graph nodes based on the centrality value in descending order. The top 20 most influential publications are listed in Table \ref{tab1} for a quick glance.

By analyzing the top 20 most influential publications above, we find that these research works can be broken down into three categories: (1) technical software programs (crystallography), (2) original synthesis experiments and (3) research reviews, which focuses on general applications and concepts. Technical software papers included those in which crystallography programs are cited. These papers are foundational in that crystal structure elucidation and validation are imperative to the any publication regarding MOFs. These included publications by Spech on single-crystal structure validation with the program PLATON and Sheldrick a short history of SHELIX. From our perspective Spech is cited more than Sheldrick because it is a historical review on SHELX, crystallographic program to solve structures, while PLATON is more impactful because it is used to validates CIF files (stands for Crystallographic Information File \cite{hall1991crystallographic}, the data exchange standard file format used in materials research) before it can be accepted into the CSD (Cambridge Structural Database, as explained in early section).

It is important to note that Table \ref{tab1} shows the reference count is not always aligned with the centrality value—for example and that the study conducted by Sheldrick is referenced significantly more than the study from Spek, but the latter study has higher centrality.

Original synthesis foundations include the preparation and discovery of new MOFs. This set includes the works by \cite{li1999design}, \cite{eddaoudi2002systematic}, and \cite{seo2000homochiral}. The work by Li presented MOF-5 and introduces the importance of the polyoxometalate clusters in the formation of strong bonds, directionality, and the use of building blocks in forming targeted frameworks. This discovery led to the first demonstration of  permanent micro porosity in MOFs and is considered the beginning of the field. Eddaoudi’s two papers introduce the isoreticular principle in the form of isorecticular expansion and functionalization, which represents the first example of chemical control over the topology of a framework. Eddaoudi also emphasizes the importance of the building blocks from a geometric perspective and the ability of the researcher to predict and thus design framework topologies. Seo work presents a homochiral MOF for enantioselective separation and catalysis which embodies the fundamentals layed forth in the topological design of Eddaoudi and solid state synthesis properties achieved by from the topology itself.

This paper is highly influential since not only does it provide the first structure property relationship paper in MOFs but also opens up the door to catalysis a property highly studied and valued property within chemistry. These four fundamental papers integrate the use of the SBUs and the isoreticular principle. This addresses the anomalies found in coordination polymers and zeolites; lack of permanent porosity in coordination polymers, lack of general functionalization in zeolites as well as the inability to predict crystal structures. This paradigm shift highlights the central theorem in MOFs the expansion of solid-state crystal engineering with encoded properties.

Reviews are broken down into either foundation (conceptual) or applications. The foundational review by Yaghi (2003) defines reticular chemistry as a logical approach “to the synthesis of robust materials with pre-designed building blocks, extended structures, and properties.” Here he proposes that MOFs can therefore be considered a subclass of crystal engineering. The review overviews the challenges from the pre-paradigm perspective by proposing two questions.

\begin{enumerate}
    \item First, of the almost unlimited possible networks, which can be expected to form and how can they be synthesized?
    \item Second, with few exceptions, MOFs based on M-N linkages in which the vertex of the network is just a single atom have a tendency to form structures which collapse upon removal of ‘guest’ atoms from the pores, rendering the structure nonporous. How then can we ensure that our structures will be robust?
\end{enumerate}

The first question answered by considering the geometries of the SBUs (secondary building units, key components in MOF structure) and linkers (connections between molecules, components in MOF structure) as well as the symmetry of the structure and the number of edges and vertices contained within the structure. This a fundamentally important concept called minimal transitivity principal and governs the default structures that are formed. The second question is discussed by overviewing rigidity and directionality of the polycarboxylate SBUs, their adopted geometries, as well as the ability of the reactants to retain structure under synthesis conditions. An important aspect of the review addressed was pore size capacity in relation to Zeolites. Overall, Yaghi discusses complexity in terms of non-default structures and provides a conceptual approach to aid the designer in MOF synthesis.

The second most impactful conceptual review is written by \cite{furukawa2013chemistry}. MOFs at this time had existed for over ten years and this review provided possible one of the second scientometric analysis of MOFs by reporting the number of MOF structures over time. The review also compared the progress in the synthesis of ultrahigh porosity of MOFs overtime. Post-synthetic modification strategies, multivariate approach, and topological design which has aided in the selectivity, complexity, and synergistic effects in MOFs were highlighted. The resulting applications from this design space such as gas storage, separation, catalysis, fuel cells, super capacitors, membrane, and thin film applications were overviewed. Thus, marking the significance and evolution of the field to that date. Other foundational reviews include Kitagawa \cite{kitagawa2004functional}, Moulton \cite{moulton2001molecules}, Ferey \cite{ferey2008hybrid}, and Batten \cite{batten1998interpenetrating}; Batten's review is on coordination polymers, Kitagawa on MOF design containing the scientometric analysis of number of publications containing coordination polymer by year from 1990 to 2002, Moulton’s on polymorphism with the distinction between crystal engineering versus crystal structure prediction, and Ferey’s distinction between SBUs in zeolites versus highly porous MOFs. These reviews provide the foundational understanding required to access various nets in MOFs. The reviews containing applications are Cui \cite{cui2012luminescent} on Luminescent functional metal organic frameworks, Li’s selective gas adsorption and separation in metal-organic frameworks \cite{li2009selective}, Lee \cite{lee2009metal} metal organic framework materials as catalysts, Sumida \cite{sumida2012carbon} carbon dioxide capture in metal organic frameworks, Li \cite{li2012metal} metal-organic frameworks for separations, Allendorf \cite{allendorf2009luminescent} luminescent metal organic frameworks, Kreno \cite{kreno2012metal} metal organic framework material as chemical sensors, and hydrogen storage in metal organic frameworks.

\begin{figure}[!h]
 \centering
 \includegraphics[height=8.5cm]{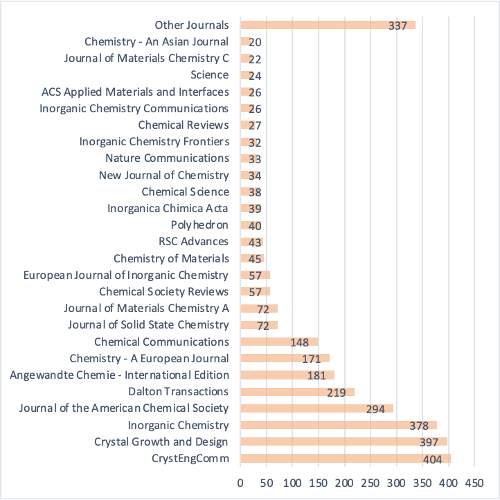}
 \caption{Journals Where Highly Impactful Research Were Published}
 \label{fig3}
\end{figure}

The journals where impactful MOF studies are published are listed in Figure \ref{fig3}. We find that most impactful journals in MOFs are domain-specific (e.g., CrysEngComm, Crystal Growth and Design), instead of high-impact journals in general (e.g., Science, Nature).

\subsection{Topic Trend Over Time}

\begin{figure*}
 \centering
 \includegraphics[height=10.5cm]{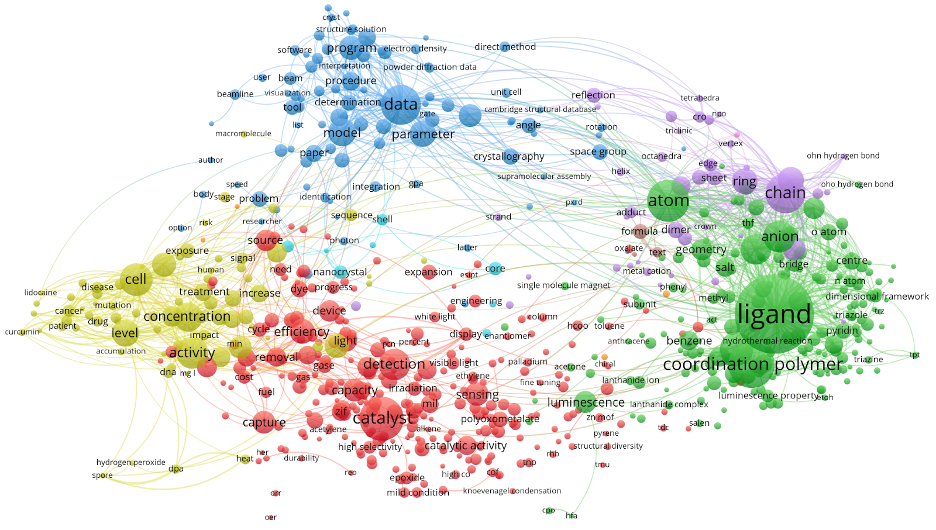}
 \caption{Keywords Extracted from MOF Research Articles and Potential Communities}
 \label{fig4}
\end{figure*}

We extracted research key terms from abstracts then constructed a keyword network. Figure \ref{fig4} is an illustration of the constructed keyword network - each node represents one term, and its size represents its occurrence. Colored nodes are used to represent articles in the same community.

Based on the keyword network above, four main communities are detected. By analyzing terms inside each community, we found that the four detected communities point to four main MOF research topics: 1) MOF synthesis, 2) properties/applications of MOF materials, 3) use of MOF in biomedicine and 4) MOF data processing and modeling. A list of example terms in each community is provided in table \ref{tab2}.

\begin{table*}[]
\centering
\caption{\ Example Terms from the Above Four Main Research Communities}
\label{tab2}
\begin{tabular}{llll}
\hline
\textbf{Properties} & \textbf{Biomedicine} & \textbf{Data Processing \& Modeling} & \textbf{Synthesis Experiment} \\
\hline
Catalyst & Cell & Data & Ligand \\
Uptake capacity & Cytotoxicity & Program & Hydrothermal condition \\
Chemical stability & Inhibitor & Processing & Single Crystal X-ray   diffraction \\
Permanent porosity & Mutation & Software & Elemental analysis \\
Durability & Protein & Interface & Topology \\
Low cost & Treatment & Model & Solvothermal reaction\\
\hline
\end{tabular}
\end{table*}

\begin{figure*}
 \centering
 \includegraphics[height=10cm]{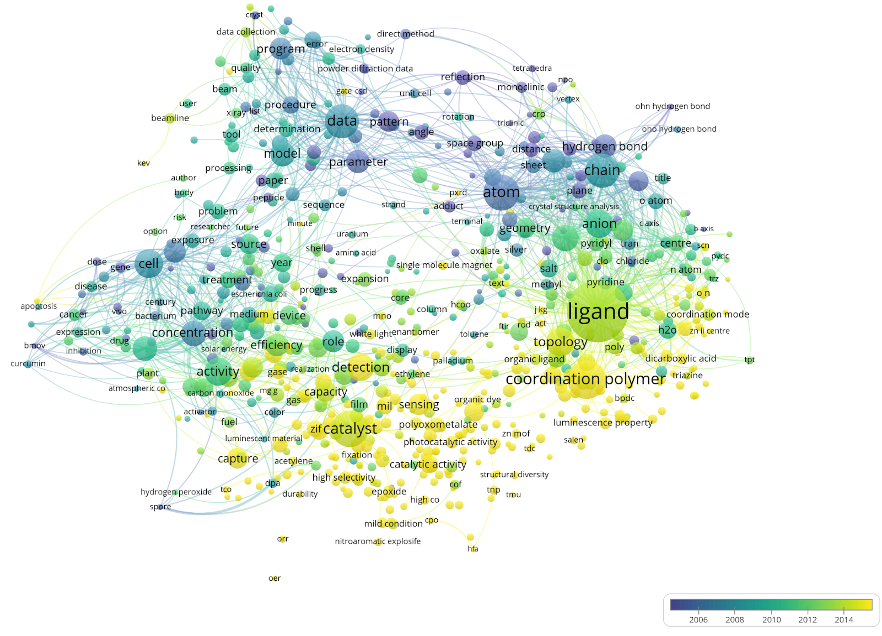}
 \caption{Trend of Research Topics Over Time}
 \label{fig5}
\end{figure*}

We analyzed the publication year of extracted research key terms to understand research trends in the MOF field. As shown in Figure \ref{fig5}, each key term in the network is colored by its average publication year. The result suggests that MOF data modeling and using MOF materials in biomedicine have attracted researchers’ attention since early years; more recently, their research trends seem to move to discover effective synthesis strategies of MOF materials, as well as the property and potential application of synthesized MOFs around the year of 2014.

\subsection{Document Co-Citation Analysis and Topic Modeling for Research Communities}

\begin{figure*}
 \centering
 \includegraphics[height=10cm]{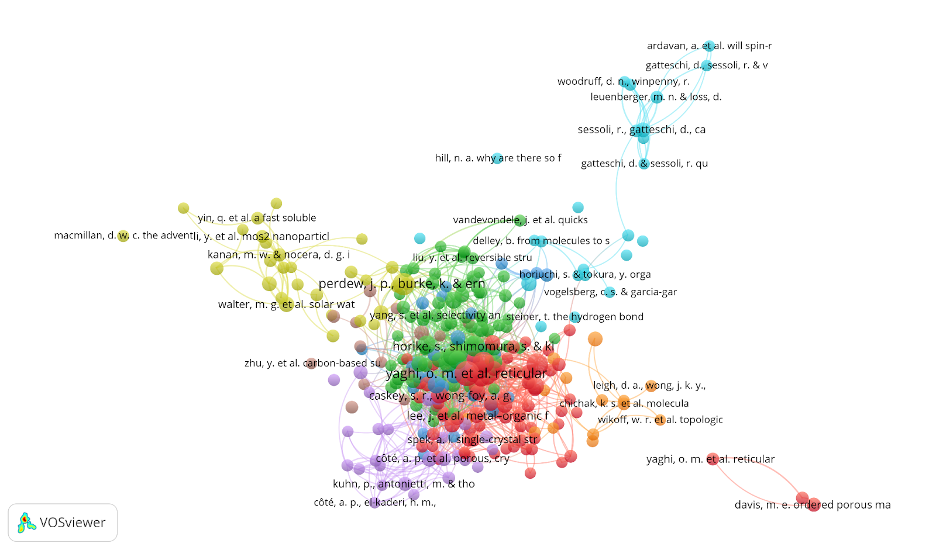}
 \caption{Potential specific Research Groups Recognized by Network Analysis}
 \label{fig6}
\end{figure*}

We used co-citation analysis to further explore specific groups in the MOF area, then used topic modeling techniques and indexed terms to unveil research interests of each specific group. To construct a co-citation network, we limited the minimum number of citations of a cited reference to 5 and kept the largest connected component. The resulting network is reported in Figure \ref{fig6}. Like the high-level topic trend analysis reported above, we detected 8 distinct specific groups in the network.

To take a closer look at specific research topics, we applied noise-reducing LDA-based topic modeling algorithms \cite{churchill2021topic} to identify topics at every specific group. Meanwhile, indexed terms in each article are also taken into account. We analyzed identified topics from each specific group. Finally, the result shows that there are at least eight specific MOF research directions. These eight specific interests can be summarized as follows: 1) drug delivery, 2) zeolitic imidazolate frameworks, 3) Carbon capture and storage, 4) Catalytic materials, 5) homochiral MOFs and asymmetric catalysis, 6) magnetic MOF materials, 7) self-assembly MOFs and 8) study on the electrical properties of MOFs. 

\section{Discussion}

As an emerging type of material, Metal-Organic Frameworks (MOFs) are garnering significant interest from the scientific community, driven by their exceptional potential and unique material characteristics. Scientometric methods play a crucial role to not only better understand the current knowledge structure, but also advance the study of metal-organic frameworks. MOFs research, which intersects chemistry, materials science, medicine and engineering, produces a vast body of literature in different scientific domains. Scientometric analysis is uniquely positioned to handle this complexity, by using data-driven quantitative methods to dissect and comprehend the extensive volume of research. 

While general methods supporting scientometric analysis, such as network analysis and NLP, can reveal important information, an exclusively data-driven approach lacks human expert feedback. Research shows that engaging domain experts provides greater insight and enriches the data analysis \cite{zanzotto2019human,wu2022survey,xin2018accelerating}. A significant contribution of the research presented in this paper is the iterative engagement of the domain experts, specifically research scientists, who helped reveal a more comprehensive view of the MOF research intellectual landscape. Human experts bring domain knowledge to the study allowing greater depth of analysis. In our analysis we integrated their knowledge about historical developments, current bottlenecks and future implications. In addition, given the complexity of scientific study such as MOFs, we found that some connections between disparate pieces of research could be overlooked by purely data-based analysis. Similarly, subtle aspects often lost when they are not immediately apparent in raw data. Domain experts involvement allowed us to mitigate the above limitations and also supported a more  in-depth interpretation on top of data analysis result by evaluating the quality and significance of specific pieces of studies. They also provided insights on how different studies, even from various disciplines, could enrich each other. Finally, human scientists can suggest innovative methodologies as well as future research direction based on the initial data analysis result, including those that are not obvious from data alone.

\section{Conclusions}

In this study, we demonstrate the value of scientometric methods enpowered by human expertise as a means for understanding the evolution of the field of MOFs. we incorporated both data and human aspects while discovering the big picture of MOFs research. We first conducted data-driven scientometric analysis, such as bibliometric, network and text analysis, then we enhanced the initial result by domain expertise from human researchers to deliver a well-presented and comprehensive intellectual landscape of metal-organic frameworks studies. 

Specifically, by integrating the result of scientometric analysis and expertise from domain scientists, (1) we identified most impactful journal, research works along with their connections in the MOF area; (2) we determined the major research directions and their trending over time by clustering algorithm; (3) we discovered specific research groups and topics in the MOF research network.

Compared to previous studies, our work greatly improved the intellectual landscape of MOFs study in the following ways: to the best of our knowledge, (1) the number of publications we collected for our analysis exceeds the volume gathered in other studies by more than tenfold, which enabled us to understand detailed landscape; (2) we introduced network analysis and natural language processing techniques to obtain data-driven exploration result, and (3) we further enhanced the data exploration result by human expertise to identify nuances overlooked by data alone.

As a future step, we plan to conduct further analysis on specific research topics we detected in this paper. For example, we are interested in extracting information from scientific literature to understand potential connections between synthesis routes and resulted material characteristics.

\bibliographystyle{unsrtnat}
\bibliography{references}  






\end{document}